# Excitonic beam steering in an active van der Waals metasurface

Melissa Li[†], Claudio U. Hail[†], Souvik Biswas, Harry A. Atwater[*]

Thomas J. Watson Laboratory of Applied Physics, California Institute of Technology,

Pasadena, California 91125, United States

*Corresponding author: Harry A. Atwater (haa@caltech.edu)

[†]-These authors contributed equally to this work

## Abstract

Two-dimensional transition metal dichalcogenides (2D TMDCs) are promising candidates for ultra-thin active nanophotonic elements due to the strong tunable excitonic resonances that dominate their optical response. Here we demonstrate dynamic beam steering by an active van der Waals metasurface that leverages large complex refractive index tunability near excitonic resonances in monolayer molybdenum diselenide ($MoSe_2$). Through varying the radiative and nonradiative rates of the excitons, we can dynamically control both the reflection amplitude and phase profiles, resulting in an excitonic phased array metasurface. Our experiments show reflected light steering to angles between -30° to 30° at three different resonant wavelengths corresponding to the A exciton, B exciton, and trion. This active van der Waals metasurface relies solely on the excitonic resonances of the monolayer $MoSe_2$ material rather than geometric resonances of patterned nanostructures, suggesting the potential to harness the tunability of excitonic resonances for wavefront shaping in emerging photonic applications.

**Introduction**

Optical metasurfaces enable powerful light manipulation by controlling the amplitude, phase, polarization, and frequency of light at the subwavelength scale with an array of point-like scatterers at an interface[1]. This principle has been used to achieve versatile wavefront manipulation on platforms such as beam deflectors[2,3], ultra-high numerical aperture lenses[4,5], and on-chip holograms[6,7]. However, due to their static nature, the functions performed by these structures are fixed at the time of fabrication. Reconfigurable functions can be achieved by introducing a material with dynamically tunable optical properties[8]. This is done in active metasurfaces, where a reconfigurable material changes the scattering properties of an array of identical metasurface unit cells. Active metasurfaces based on co-integration of scatterers and active materials have been realized with transparent conducting oxides[9–11], liquid crystals[12], phase-change materials[13–15], thermo-optic materials[16] and graphene[17], which led to demonstrations of dynamic beam scanning[9,18,19], varifocal lenses[20], and polarization conversion[21]. However, these concepts rely on coupling tunable materials with resonant nanostructures, cavities, or antennas that are highly sensitive to geometric parameters and typically only operate over a limited wavelength range. As structures for controlling the propagation of light become increasingly complex and compact, the need for new materials that can exhibit unique, strong light-matter interactions in the ultra-thin limit is growing rapidly.

Excitonic materials, such as two-dimensional (2D) van der Waals materials, are especially attractive as their resonant properties are intrinsic to the material and do not rely on coupling to cavities or antennas[22,23]. In the monolayer limit, the optical response of 2D transition metal dichalcogenides (TMDCs) is dominated by excitonic resonances[22,24–26], which are highly tunable with external electric field or doping[27,28], dielectric environment engineering[29,30], and strain[31,32].

The strong light-matter interactions from these excitonic resonances provide opportunities to design active optical devices where a single monolayer TMDC can act as both the resonant scatterer and the tunable material. Among different TMDCs, molybdenum diselenide ($MoSe_2$) exhibits especially large tunability of its optical properties at cryogenic temperatures. Without external enhancement or coupling to optical cavities, excitonic reflectivity of over 80%[33,34] and tunability of over 200% in the real and imaginary part of the refractive index[35] have been shown by modifying the radiative and nonradiative rates in $MoSe_2$. These rapid advances in understanding exciton resonances have stimulated thinking about active metasurfaces that exploit the excitonic modulation phenomena to enable van der Waals active metasurfaces[36,37]. For example, excitonic resonances have been used to demonstrate an intensity-tunable lens with a fixed focal length[38] or to steer a focused beam at an edge of three $MoSe_2$ flakes[39]. However, despite much progress, the dynamic control thus far has been limited to amplitude modulation or local point control of an edge, and versatile wavefront control of reflected light from a single atomic layer has remained an outstanding challenge.

Here, we show that tunable exciton resonances in $MoSe_2$ can be harnessed to design an excitonic phased array metasurface for dynamic beam steering. We can model the gate-tunable refractive index by modifying the radiative and nonradiative rates of the excitons with applied potential, which gives rise to large amplitude and phase modulation of the scattered light. We also experimentally show how under varying voltage configurations, the reflected light can be steered to different angles at three wavelengths corresponding to the A exciton, B exciton, and trion resonances. The observation of the beam steering at three distinct excitonic wavelengths in the absence of coupling to antennas or gratings reveals the steering is independent of geometric resonances. Our results can be extended to other classes of ultra-thin materials that exhibit tunable excitonic resonances such as black phosphorus[36,40], semiconductor quantum dots[41,42], and layered perovskites[43,44], opening a new path for designing atomically-thin, reconfigurable metasurfaces.

## Results & Discussion

Figure 1 illustrates an atomically thin, excitonic metasurface consisting of a monolayer of $MoSe_2$ encapsulated in hexagonal boron nitride (hBN) on subwavelength-spaced, lithographically patterned gold electrodes, supported on a thermally oxidized silicon substrate. Light impinges at normal incidence polarized along the y direction, and is reflected after passing through the heterostructure. The refractive index in the $MoSe_2$ layer is periodically modulated along the x direction via a gate voltage V(x) applied to the periodically arranged gold electrodes. This local modulation imparts an electrically tunable phase shift φ(x) to the reflected wavefront. Imposing a constant phase gradient on the reflected wavefront then deflects the reflected beam to an angle. Figure 1b illustrates this electrically controlled light deflection at the A exciton wavelength with the electric field intensity in a cross-section of the surface as calculated by full wave electromagnetic simulations.

In designing our metasurface, we first consider the voltage dependent optical response of monolayer $MoSe_2$ at 6 K. At cryogenic temperatures, the optical response of $MoSe_2$ is dominated by the A exciton, trion, and B exciton resonances where the linewidths of these three oscillators are much narrower than their linewidths at room temperature[33]. By exploiting the narrow excitonic linewidths at low temperatures, combined with Fermi level control via electrical gating of the carrier density, the refractive index of $MoSe_2$ can be modulated by over 200%[35]. This tunability results from the interplay between radiative and nonradiative decay channels with gating. The dielectric function of $MoSe_2$ takes the form of a sum of Lorentz oscillators[33,45]:

$$\epsilon = \epsilon_1 + i\epsilon_2 = \epsilon_\infty - \frac{\hbar c}{d_{MoSe_2}} \sum_j \frac{\hbar \gamma_{r_j}}{E_j(E - E_j + i\hbar \gamma_{nr_j}/2)}.$$

Here, $\epsilon_\infty$ is the background MoSe$_2$ dielectric constant, $d_{MoSe_2}$ is the thickness of the monolayer, and $E_j$, $\gamma_{r_j}$, and $\gamma_{nr_j}$ are the resonant energy, radiative, and nonradiative emission rates of the j[th] oscillator, respectively. In our analysis, we model the dielectric function with two oscillators corresponding to the A exciton and trion and add a background dielectric constant, $\epsilon_\infty = 21$, to account for oscillators corresponding to higher energy optical resonances of the material outside of our measurement window[46]. The refractive index of MoSe$_2$ is then obtained by $\tilde{n} = n + ik = \sqrt{\epsilon}$. The photoluminescence quantum yield is defined as $\eta = \gamma_r / (\gamma_r + \gamma_{nr})$, where the rates correspond to the intrinsic radiative and nonradiative emission rates of the A exciton at the charge neutral point. We set the charge neutral point to be at -5 V since our MoSe$_2$ sample is n-doped. Electrical gating of the carrier density then modifies the radiative and non-radiative emission rates, yielding the gate tunable refractive index (see Fig. S1, S2 and Supplementary Note 1). Depending on $\eta$, the MoSe$_2$ dielectric function takes a negative or positive value on resonance, yielding an epsilon-near-zero (ENZ) response and hence a tunable transition from metallic to dielectric near the exciton energy, resulting in large changes in reflectance amplitude and phase.

Figures 2a and b illustrate the simulated gate voltage dependent reflection amplitude and phase of our atomically thin metasurface. Here we choose a $\eta = 0.5$, and a top and bottom hBN thickness of 6 nm and 60 nm, respectively, with gold electrodes that are 20 nm thick, 380 nm wide and spaced at 120 nm apart (simulated reflectance and phase spectrum for $\eta = 0.25$ are shown in Fig. S3). The reflection spectrum is dominated by the gate-tunable A exciton resonance and thin film interference from the substrate and heterostructure. With the incident light polarized along the gold contacts, tunable reflection originates only from the A exciton of MoSe$_2$, as the local surface plasmon mode of the electrodes is not excited (the corresponding analysis and discussion for illumination with the perpendicular polarization are shown in Fig. S4 and Supplementary Note 2). The gate voltage induced reflected light phase shift is maximized near the A exciton resonance wavelength of 757 nm, and reaches up to a range of 240°. Figure 2c and d show the voltage dependent amplitude and phase of the reflected light at the A exciton resonance for $\eta = 0.25$, 0.5 and 0.75. Notably, the available gate voltage-induced phase modulation range strongly depends on the quantum yield of the $MoSe_2$, with a higher $\eta$ resulting in an increased tuning range. While large values of $\eta$ of up to 90% have been reported, it is important to note that there can be a significant variation between different monolayer flakes, and also variation within the same flake due to charge and strain inhomogeneities[34]. Hence, for larger area structures, such as the one proposed here, the effective averaged quantum yield is reduced compared to the reported peak values. However, even with moderate quantum yields ranging between 25-50% a significant phase modulation range is attained. Setting the gate voltage along the metasurface to produce a constant phase gradient (according to the look up table in Fig. 2d), allows steering the reflected beam to arbitrary angles θ. Figure 2e and f show the far field intensity pattern of seven different periodic voltage configurations $V(x)$ that steer the reflected beam between –30° to 30° for $\eta = 0.25$ and $\eta = 0.5$. The seven different voltage configurations generate a phase gradient as required for deflecting the light to the specified angle. The phase periodicities for the angles 30°, 22° and 18° are 1.5, 2 and 2.5 μm, respectively, where the periodicity was varied by changing the number of electrode elements in each period. Even with a reduced quantum yield, light can be deflected to varying angles, although the specular reflection remains intense. With a quantum yield of 50%, most light is steered into the desired direction, and relatively little intensity is specularly reflected or coupled to the opposite direction.

Figure 3a shows a microscope image of the fabricated metasurface, which contains both a monolayer (blue) and bilayer (red) portion of a MoSe$_2$ flake encapsulated with hBN. We fabricated the metasurfaces by dry transfer of an hBN/MoSe$_2$/hBN heterostructure onto a lithographically patterned substrate with gold electrodes (see methods for details). The electrodes were patterned to allow setting a periodic three level voltage gradient, with the first set connected to an external voltage $V_1$ and the third set to $V_3$, and the second set of electrodes floating. The fabricated surfaces were characterized in a cryostat at a temperature of 6 K by performing widefield reflectance microscopy with a coherent broadband illumination (see methods for details). Figure 3b shows a spatial reflectance map of the ungated metasurface near the A exciton resonance at 757 nm. Figure 3c illustrates reflectance spectra measured on the monolayer and bilayer region at the positions marked in Fig. 3b. In the monolayer region, a clear reflectance minimum is observed at 757 nm as expected due to the A exciton. We note the exciton in the bilayer is redshifted to 767 nm from interlayer hybridization effects and is broader than the exciton in the monolayers[47].

To characterize the beam steering, we performed spectrally selective Fourier plane imaging using reflected light. For this measurement, we illuminate a large area of the surface (10 μm in diameter) with a near collimated beam. To avoid effects arising from surface nonuniformity, we limit light collection to an area of 4 μm in diameter, by means of spatial filtering with an aperture in the image plane (see Fig. S5). Figure 3d illustrates the measured spectral Fourier plane image on the monolayer portion of the sample for four different voltage configurations. For a constant voltage, $V_1 = V_3 = 0$ V, the incident beam is specularly reflected to $\theta = 0°$, with some light coupled to symmetric side lobes at $\theta = \pm15°$ that arise from the finite imaging aperture. Imposing a symmetrically varying voltage profile, $V_1 = V_3 = -5$ V, on the surface couples light into the +1 and –1 diffraction order at $\theta = \pm30°$ at the A exciton wavelength 757 nm. Notably, the two deflected beams are separate from the specular reflection and close to collimated, with an angular divergence of 16°. This is a stark contrast to previous demonstrations, where the deflected beam was highly divergent[39]. The asymmetry in the intensity of the +1 and –1 diffraction order is due to the finite size aperture used in the detection. By applying a voltage gradient to impose a linear phase gradient along the interface, $V_1$ = 5 V and $V_3$ = -5 V, we show the asymmetric steering of light into the +1 diffraction order, and the suppression of the –1 order. Similarly, by applying an inverted voltage gradient, $V_1$ = -5 V and $V_3$ = 5 V, we steer light into the –1 diffraction order and suppress the +1 order. In all three cases, the beam steering occurs selectively at the A exciton wavelength, which is further evidence of the exciton-induced tunability. The small asymmetry in steering light to the +1 and –1 is attributed to the finite sized aperture in the detection. Figure 3f shows a spatial map of the measured efficiency of coupling reflected light into the +1 diffraction order at various locations of the surface with $V_1$ = 5 V and $V_3$ = -5 V. On the monolayer flake, outlined in blue, there is strong spatial variation in the diffraction efficiency, which we attribute to the spatial inhomogeneity of the linewidth and quantum yield. The region with the highest diffraction efficiency is expected to concur with the region of highest quantum yield in the $MoSe_2$. In the bilayer region, the diffraction efficiency vanishes at 757 nm, but is observed at the redshifted bilayer exciton resonant wavelength of 767 nm (Fig. S6).

Next, we study the voltage dependence and spectral dependence for the different diffractive orders at the A exciton energy. In Figure 4a and 4b, we see the intensity at 30° is maximized near the charge neutral point of the $MoSe_2$ between -4 V to -5 V, where the A exciton linewidth is the narrowest. As the carrier density increases through either hole or electron doping, the intensity at 30° decreases due to lower oscillator strength and linewidth broadening from increase in nonradiative rates. We also observe the blueshift in A exciton resonance energy, which is consistent with previously observed blueshift by Pauli blocking upon doping. The gate dependent diffraction efficiencies for the -1, 0 and +1 diffraction order with the corresponding simulation results are shown in Figures 4c, d, and e, respectively. Under the voltage gradient profile, we see an exchange of the diffraction efficiency of the $0^{th}$ order and +1 order, whereas the diffraction efficiency at the -1 order shows minimal dependence on the applied voltage. At -4 V, the diffraction efficiency is in good agreement with our simulated result of an averaged $\eta = 0.25$ (Fig. 4c-e and Fig. S7), which confirms the observed beam deflection arises from excitonic phase modulation in $MoSe_2$.

Aside from the A exciton resonance at 757 nm, we also observe resonances corresponding to the B exciton and trion in $MoSe_2$. The spectral Fourier plane images near the B exciton wavelength are shown in Figure 5a. For $V_1 = V_3 = 5$ V, we see minimal beam deflection to the +1 order (26°) due to the suppression of the B exciton resonance. However, under an applied voltage gradient, with $V_1 = 5$ V and $V_3 = -5$ V, we observe an enhancement in the intensity at the +1 diffractive order. Figure 5b shows the voltage ($V_3$) dependent diffraction efficiency at the +1 order for wavelengths between 650 nm to 800 nm, where $V_1 = 5$ V and $V_3$ is varied. We observe three prominent features at 670 nm, 757 nm, and 770 nm that correspond to the resonances of the B exciton, A exciton, and trion, respectively, suggesting that the incident beam can be steered to a desired angle for three different excitonic resonant wavelengths. Similar to the A exciton, the B exciton diffraction efficiency is also maximized near the charge neutral point. However, the deflection efficiency is lower for the B exciton since the linewidth is broader than the A exciton, resulting in lower refractive index modulation and phase shift. Moreover, the lower phase shift also results in less suppression of the -1 order, so the beam is deflected to both the -1 and +1 order even under an asymmetric voltage gradient (Figure 5c-d). Nevertheless, the observation of steering at three different wavelengths that correspond to excitonic resonances shows our design solely relies on tunable excitonic resonances in MoSe$_2$ -a material property- and not geometric antenna or grating resonances.

**Conclusions**

We employ tunability in the complex refractive index of monolayer MoSe$_2$ at the A exciton, B exciton, and trion to enable multi-wavelength excitonic beam steering at the nanoscale. We find the steering efficiency and directivity are limited by the homogeneity in the radiative efficiency over the metasurface aperture, here comprising a few microns across the flake. While our demonstration of excitonic beam steering was performed on a 15 μm flake at 6 K, large aperture area room temperature metasurface operation could potentially be achieved through methods of gold assisted exfoliation[48,49] and through suppression of nonradiative decay mechanisms at higher temperatures, such as superacid treatments[50] or strain[51]. Advances in large area fabrication of 2D materials would enable more complex device operations, such as multifunctional excitonic metasurface that demonstrate both tunable focal length and beam steering, or space-time modulated metasurface based on the nanosecond response time of excitons. Our findings provide a route for exploring arbitrary wavefront shaping based on the tunability of excitons.

## Methods

**Measurement** Low temperature optical spectroscopy was performed using an attoDRY800 system, which allows for temperature control. The sample was mounted on a piezo-stage using Apiezon thermal grease and the stage was cooled by closed-cycle circulating liquid helium. For the optical characterization, broadband light from a supercontinuum laser (NKT SuperK Extreme) was impinged on the sample in a Köhler illumination configuration passing through a vacuum and low temperature compatible objective (NA = 0.82). The reflected light was collected by the same objective and directed to a CCD camera and a grating spectrometer (See Fig. S5). For Fourier plane imaging, a Fourier plane was formed at the entrance slit of the spectrometer through a set of lenses.

**Fabrication** Samples were fabricated by mechanically exfoliating hBN and monolayer $MoSe_2$ from bulk crystals (hq-graphene and 2D semiconductors) onto polydimethylsiloxane (PDMS) stamps using blue Nitto tape. The $hBN/MoSe_2/hBN$ heterostructure was then sequentially transferred from the PDMS to the patterned Au electrode array using the all-dry transfer technique. The Au electrode array and contact pads were fabricated with a sequential multistep process of electron beam lithography, evaporation, and lift-off, to allow for different thicknesses of the electrode array and electric contact pads. To improve the adhesion of Au on $SiO_2$, a 2 nm thick Ti adhesion layer was used.

**Simulations** The numerical modeling of the nanostructures was carried out using a finite difference time domain (FDTD) method with the commercially available software *Lumerical*. The gate dependent reflection, phase and beam deflection (Fig. 2) was simulated using periodic boundary conditions and a spatially coherent plane wave illumination. For the comparison between simulations and experiment (Fig. 4), a gaussian illumination with a 24 μm beam diameter and perfectly matched layer boundary conditions were used. The effect of the numerical aperture of the imaging objective lens and the finite imaging aperture was accounted for through a set of Fourier transforms and spatial filtering in the Fourier and image plane. A constant refractive index $n = 1.45$ was used for thermal oxide and $n = 2.2$ for hBN. The smallest mesh refinement of 0.2 nm was used.

## Acknowledgements

C.U.H. acknowledges support from the Swiss National Science Foundation through the Early Postdoc Mobility Fellowship grant #P2EZP2_191880. We gratefully acknowledge the critical support and infrastructure provided for this work by The Kavli Nanoscience Institute at Caltech.

## Contributions

M.L, C.U.H., and H.A.A conceived the project. M.L. and C.U.H. fabricated the samples with support from S.B.. M.L. and C.U.H. worked on measurements, performed FDTD simulations, and wrote the manuscript. M.L. analyzed the results with assistance from C.U.H.. C.U.H. built the experimental set-up. H.A.A. supervised the project. All authors discussed the implications of the results and provided important feedback.

## Competing interests

The authors declare no competing interests.

## Correspondence and requests for materials should be addressed to H.A.A.

## Data availability

The data that support the plots within this paper and other findings of this study are available from the corresponding authors upon reasonable request.

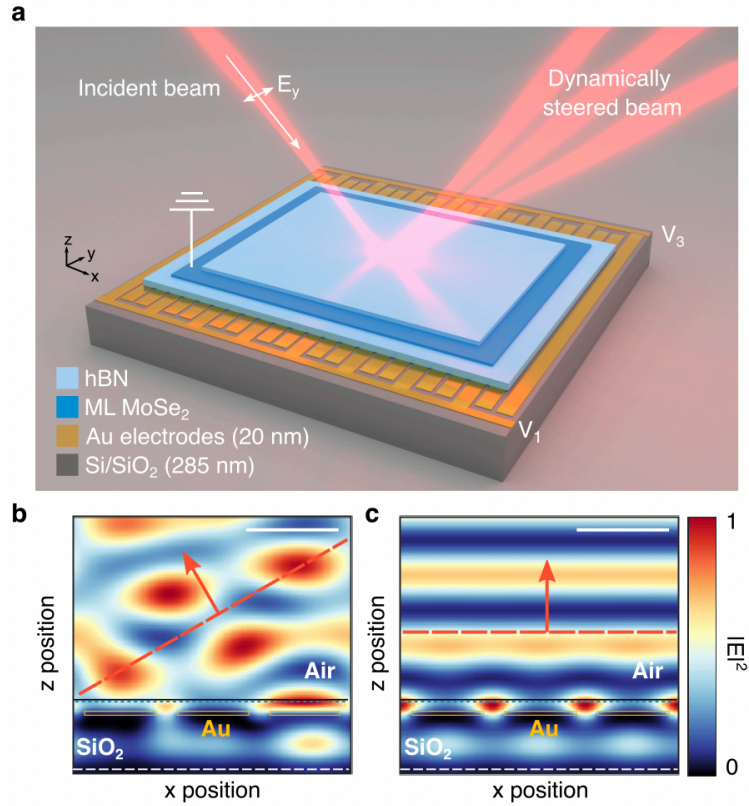

**Figure 1. | Dynamic beam deflection with a monolayer MoSe₂.** (a) Schematic of the exciton based TMDC metasurface for dynamic beam steering. (b) Simulated electric field intensity cross section at the A exciton resonance of 757 nm under an applied voltage gradient. Top hBN and air interface shown in solid black line. MoSe₂ position shown in dotted blue line. Red arrow shows the reflected beam is steered. (c) Simulated electric field intensity cross section at 757 nm under a constant applied voltage. Red arrow shows the beam is specularly reflected. Scale bars: 500 nm.

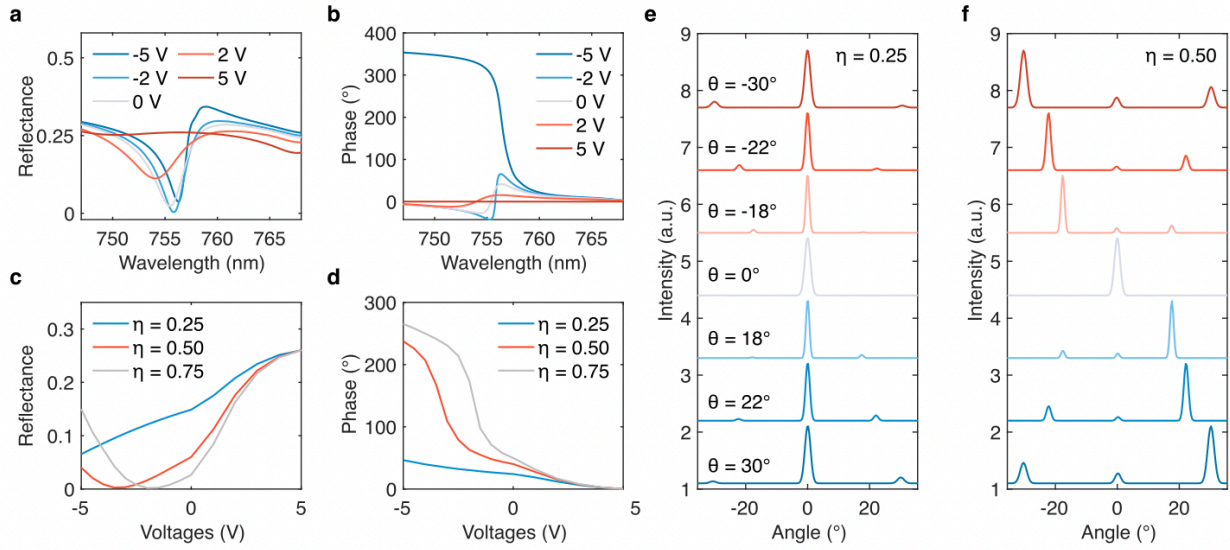

**Figure 2. | Simulated voltage-dependent reflectance amplitude, phase, and beam steering.** (a) Reflectance spectra and (b) phase for different voltages for an MoSe₂ sample with photoluminescence quantum yield η = 0.5. (c) Voltage-dependent reflectance and (d) phase at the A exciton wavelength of 757 nm for different radiative efficiencies. (e,f) Simulated reflected angular far field intensity for η = 0.25 and η = 0.5, respectively, under different applied voltage gradient profiles and corresponding steered angle.

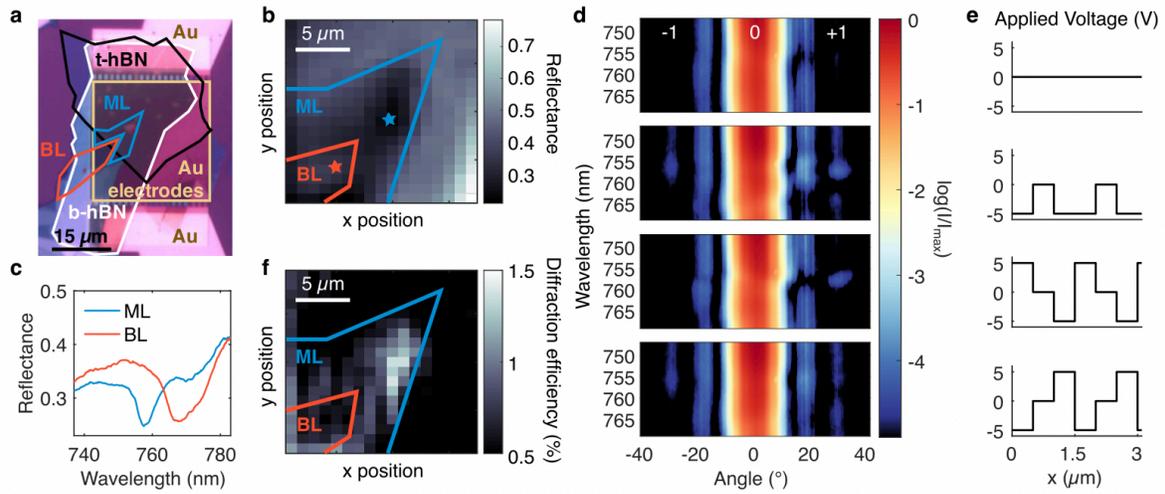

**Figure 3. | Fabrication and experimentally measured dynamic beam steering in monolayer MoSe₂.** (a) Optical microscope image of the metasurface device. (b) Spatial reflectance map at 757 nm. (c) Reflectance spectra of the monolayer (blue) and bilayer (red) MoSe₂ region. (d,e) Measured spectral Fourier plane images of the monolayer MoSe₂ metasurface near the A exciton resonance under different applied voltage configurations to each electrode for dynamic beam deflection at ±30°. (f) Spatial map of the intensity of the steered light at the +1 diffraction order (+30°). In a, b, and f, the solid blue line outlines the monolayer region (ML) and the solid red line outlines the bilayer region (BL).

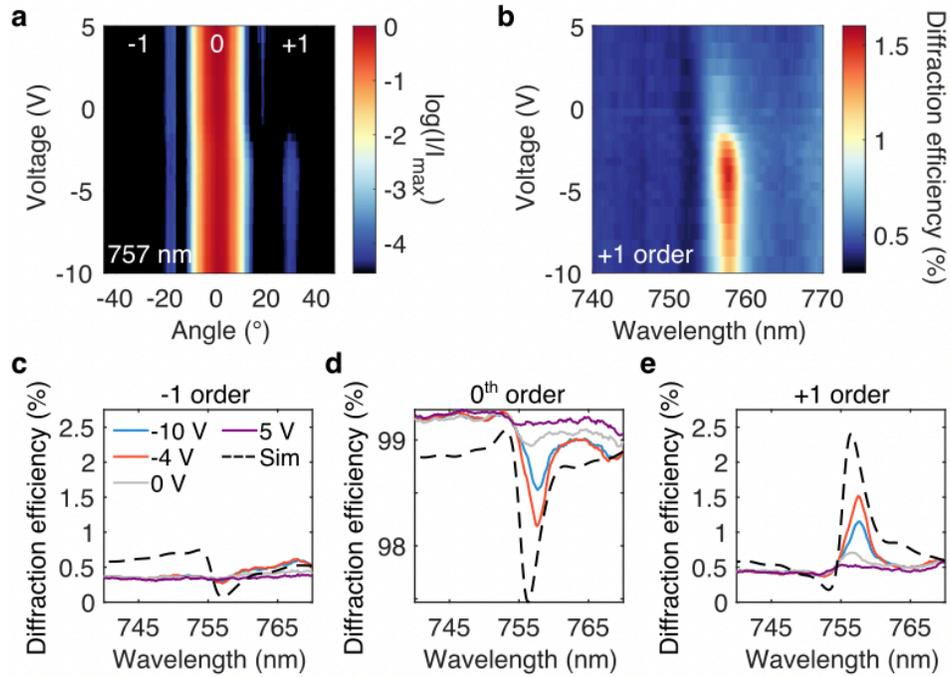

**Figure 4. | Voltage dependence of diffraction efficiency near the A exciton resonance.** (a) Measured intensity in the Fourier plane for $V_1$ = 5 V and varying voltage $V_3$ at the resonant A exciton wavelength of 757 nm. Under the applied asymmetric voltage gradient, only the +1 diffractive order is observed. (b) Voltage and spectral dependence of the diffraction efficiency of the +1 order. The intensity is maximized near the charge neutral point of the $MoSe_2$ between -4 V to -5 V and blueshifts with increase in carrier density. (c-d) Measured (solid) and simulated (dashed) diffraction efficiency for the -1, 0, and +1 order, respectively. The intensity at the $0^{th}$ order is primarily exchanged with the +1 order at the A exciton resonance.

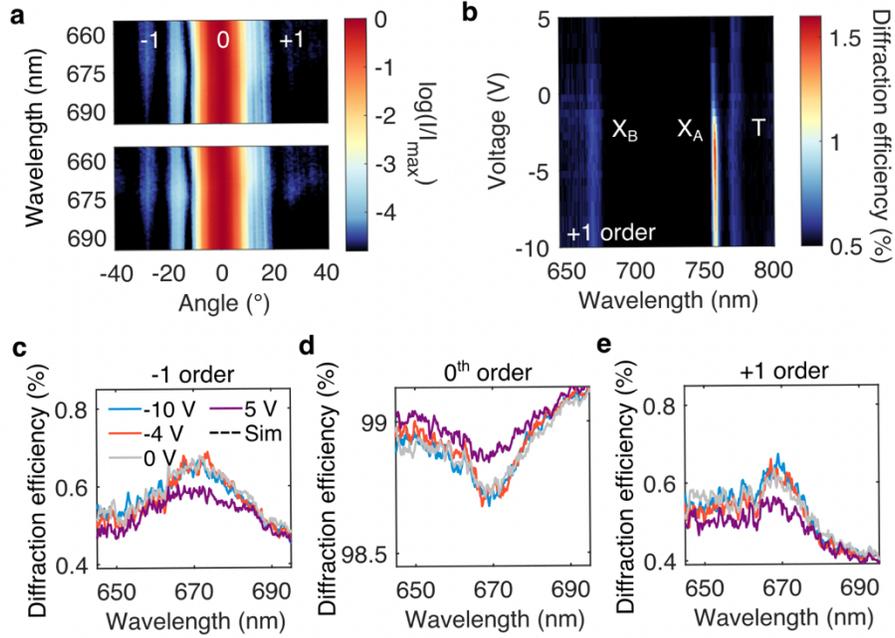

**Figure 5. | Voltage dependence of diffraction efficiency near the B exciton resonance**. (a) Measured spectral Fourier plane images of the monolayer MoSe$_2$ metasurface near the B exciton resonance. (b) Voltage and spectral dependence of the diffraction efficiency of the +1 order between 650 nm and 800 nm. Three distinct peaks corresponding to the B exciton (X$_B$), A exciton (X$_A$), and trion (T) are observed. (c-d) Measured diffraction efficiency for the -1, 0, and +1 order, respectively. The intensity at the 0$^{th}$ order is exchanged with both the -1 order and +1 order at the B exciton resonance.